\documentclass[10pt,letterpaper]{article}
\usepackage{opex3}
\usepackage{cite}

\begin{document}
\title{Tuning the phase sensitivity of a double-lambda system with a static magnetic field}
\author{Xiwei Xu, Shuo Shen and Yanhong Xiao}
\address{Department of Physics, State Key Laboratory of Surface Physics, and Key Laboratory of Micro and Nano Photonic Structures (Ministry of Education), Fudan University, Shanghai 200433, China}
\email{yxiao@fudan.edu.cn}
\date{\today}

\begin{abstract}
We study the effect of a DC magnetic field on the phase sensitivity of a double-lambda system coupled by two laser fields, a probe and a pump. It is demonstrated that the gain and the refractive index of the probe can be controlled by either the magnetic field or the relative phase between the two laser fields. More interestingly, when the system reduces to a single-lambda system, turning on the magnetic field transforms the system from a phase-insensitive process to a phase-sensitive one. In the pulsed-probe regime, we observed switching between slow and fast light when the magnetic field or the relative phase was adjusted. Experiments using a coated $^{87}$Rb vapor cell produced results in good agreement with our numerical simulation. This work provides a novel and simple means to manipulate phase sensitive electromagnetically-induced-transparency or four-wave mixing, and could be useful for applications in quantum optics, nonlinear optics and magnetometery based on such systems.
\end{abstract}



\section{Introduction}

Quantum interference between different excitation path ways is an intriguing phenomenon in quantum optics, laser and atomic physics. Notable examples are electromagnetically induced transparency (EIT) \cite{1EIT-Harris,2EIT-lukin,3EIT-Fleischhauer}, multi-wave mixing \cite{4multi-wavemix} etc. Among these, an interesting category is the phase-sensitive process formed by closed-loop interactions, where the phases of the optical fields can dramatically change the steady state of the atoms and the optical susceptibility. There has been a large amount of work in this field since the 1980s \cite{5-1expforPhaseSensitive,5-2expforPhaseSensitive}, and the interests boosted after the experimental demonstration of phase-sensitive EIT \cite{15PhaseDep}. Applications of phase sensitive processes include large nonlinearity \cite{6largenonlinearity,7largenonlinearity,8largenonlinearity}, slow and fast light \cite{9fastslowlight,10fastslowlight,11fastslowlight}, and optical switch \cite{12OptSwitch} etc. To the best of our knowledge, it is a tradition to use oscillating electromagnetic fields, i.e., either all optical fields \cite{15PhaseDep,13PhaseDep,14PhaseDep,16PhaseDep,17PhaseDep}, or a combination of microwave fields and optical fields \cite{18PhaseMicro,19PhaseMicro,19-1PhaseMicro}, to form a closed loop.

In this paper, we propose to use a static magnetic field to close the interaction loop. This approach is suitable as long as the ground states can be coupled by the magnetic field. For example, ground states formed by superpositions of Zeeman sublevels can be mixed if their spin orientations are not parallel to the external magnetic field. In this case, the magnetic field can in effect coherently couple the two ground states through Larmor precession. For instance, as we will show below, in a lambda system, without the magnetic field the system is a normal EIT configuration, whose steady state susceptibility is not influenced by the optical phases; with the magnetic field, a closed loop is formed and the field absorption becomes dependent on the relative phase. Similar idea also applies to a double-lambda system. Although its level configuration naturally allows a closed-cycle four-wave-mixing (FWM) which is already phase sensitive, applying a magnetic field can still alter the degree of phase sensitivity, as also shown below both theoretically and experimentally.

Adding a magnetic field as a new knob to the double-lambda system with Zeeman sublevel ground states might be particularly useful, since such a system has attracted considerable amount of interests recently for use in light squeezing \cite{20Squeeze,21Squeeze}, light entanglement \cite{22Entangle} and amplified slow and stored light \cite{23AmpSlowLight}.  The following advantages of this double-lambda system are probably responsible for its popularity: (1) It corresponds to the energy configuration of the D1 line of the alkali atoms, and can be easily implemented by experiments. (2) Its relevance to magnetic fields allows various magnetometer schemes to be realized \cite{24Magnetometer,25Magnetometer}. (3) Laser fields with only one frequency are sufficient to address this system due to the Zeeman ground states, which greatly simplifies experiments. (4) The long lifetime of the ground states enables coherence-enhanced nonlinear optics \cite{3EIT-Fleischhauer}.

This paper is organized as follows. In section 2, we introduce the system theoretically, and lay out key steps in our numerical simulation. Then in section 3, we describe our experimental setup and measurement methods. In section 4, we present the experiment results along with corresponding numerical results, which mainly include: (i) The presence of a magnetic field increases the dependence of the optical absorption on the relative phase between laser fields; (ii) When the magnetic field reverses and the relative phase changes by $\pi$, the optical responses remain the same; (iii) In the dynamical regime, the probe pulse can experience either slow light or fast light depending on the magnetic field and the relative phase. Finally in section 5, we conclude.

\section{Theory}

\begin{figure}[t]
  \centering\includegraphics[width=0.8\textwidth]{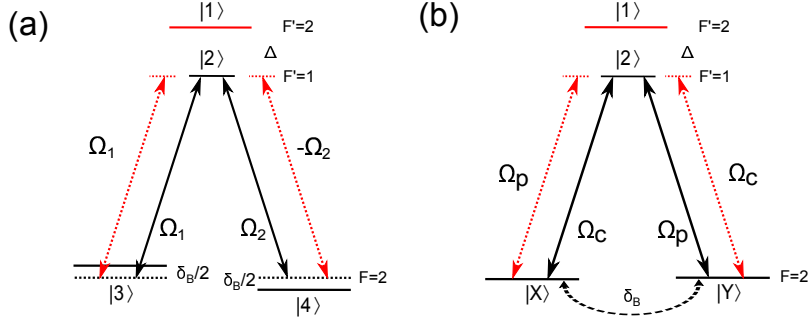}
 \caption{Level configurations for a double-lambda system coupled by two orthogonally polarized laser fields resonant with the lower excited state, represented (a) in the circular basis where the two circularly polarized fields are $E_{1}$ and $E_{2}$, and (b) in the linear basis where the two linearly polarized fields are $E_{c}$ and $E_{p}$. Here, $|X\rangle=(|3\rangle+|4\rangle)/\sqrt{2}$, $|Y\rangle=(|3\rangle-|4\rangle)/\sqrt{2}$. Due to the selection rules and C-G coefficients of Rb, $E_{1}$ couples $|3\rangle\rightarrow|2\rangle, |1\rangle$ both with Rabi frequency $\Omega_1$, and $E_{2}$ couples $|4\rangle\rightarrow|2\rangle, |1\rangle$ with Rabi frequency $\Omega_2$, $-\Omega_2$ respectively. Similarly, $E_{p}$ couples
 $|Y\rangle\rightarrow|2\rangle, |X\rangle\rightarrow|1\rangle$ both with Rabi frequency $\Omega_p$, and $E_{c}$ couples $|X\rangle\rightarrow|2\rangle, |Y\rangle\rightarrow|1\rangle$ both with Rabi frequency $\Omega_c$.}
\label{FWMpic}
\end{figure}

The double-lambda system under consideration is shown in Fig.~\ref{FWMpic}, where (a) and (b) use different ground state basis. $|3\rangle$ and $|4\rangle$ are degenerate Zeeman sublevels (with opposite magnetic quantum numbers) and are eigenstates of the atomic angular momentum component along the direction of the static magnetic field.   $|1\rangle$ and $|2\rangle$ are the two hyperfine excited states separated by $\Delta$. The optical fields are nearly resonant with the lower excited state. We assume that the selection rules and C-G coefficients determine that a sigma plus (minus) field $E_1$ ($E_2$) can couple $|3\rangle$ ($|4\rangle$) to both $|1\rangle$ and $|2\rangle$, with Rabi frequency $\Omega_{1}$ ($-\Omega_2$) and $\Omega_{1}$ ($\Omega_{2}$) respectively. Since in the experiment we use two orthogonally linearly polarized laser fields, a weak probe $E_p$ and a strong control $E_c$, it is more convenient to use the linear basis in Fig.~\ref{FWMpic} (b), where $|X\rangle=(|3\rangle+|4\rangle)/\sqrt{2}$, and $|Y\rangle=(|3\rangle-|4\rangle)/\sqrt{2}$. In this basis, the probe field couples $|Y\rangle\rightarrow|2\rangle$ and $|X\rangle\rightarrow|1\rangle$ transitions with the same Rabi frequency $\Omega_{p}=(\Omega_{1}-\Omega_{2})/\sqrt{2}$, and the control field couples $|X\rangle\rightarrow|2\rangle$ and $|Y\rangle\rightarrow|1\rangle$ transitions with the same Rabi frequency $\Omega_{c}=(\Omega_{1}+\Omega_{2})/\sqrt{2}$. When a magnetic field along the light propagation direction is applied, $|3\rangle$ and $|4\rangle$ have zeeman shift $\delta_{B}/2$, $-\delta_{B}/2$ respectively; and in the linear basis, $|X\rangle$ and $|Y\rangle$ are coherently coupled by the magnetic field with an effective Rabi frequency equal to $\delta_{B}$. Here, $\delta_{B}=g \mu_{B} m B/h$ with $\mu_{B}$ the Bohr magneton, $g$ the Land\'{e} $g$ factor, and $m$ the magnetic quantum number.

The Hamiltonian of the system takes the following form in the circular basis (Fig.~\ref{FWMpic} (a)):
\begin{eqnarray}
&\hat{H}=
2\pi\delta_{B}(|3\rangle\langle3|-|4\rangle\langle4|)/2+\Delta|1\rangle\langle1|
+\Omega_{1}|2\rangle\langle3|+\Omega_{2}|2\rangle\langle4|
\label{Hamiltonian}\\
&+\Omega_{1}|1\rangle\langle3|-\Omega_{2}|1\rangle\langle4|+H.c.\nonumber.
\end{eqnarray}
Here, $\Omega_{1}=\wp_{32}E_{1}/\hbar$ and $\Omega_{2}=\wp_{42}E_{2}/\hbar$, with $\wp_{ij}$ the transition dipole moment between level $i$ and $j$. The two circularly polarized laser fields $E_{1}$ and $E_{2}$ are connected to the linearly polarized fields $E_{p}$ and $E_{c}$ by
\begin{equation}
\left(\begin{array}{c} E_{1} \\E_{2} \\\end{array}\right)=\frac{1}{\sqrt{2}}\left(\begin{array}{cc}1 & i \\1 & -i \\\end{array}\right) \left(\begin{array}{c}E_{c}\\E_{p}\\\end{array}\right)
\label{LinearToCircular}
\end{equation}
Various decay channels can be taken into account by the master equation
\begin{equation}
\dot{\hat{\rho}}=\frac{1}{i\hbar}[\hat{H},\hat{\rho}]+\hat{\Gamma}_{exc}-\hat{\Gamma}_{rel}
\label{MasterEquation}
\end{equation}
where $\hat{\Gamma}_{exc}$ and $\hat{\Gamma}_{rel}$ are excitation matrix and relaxation matrix.
We assume that both excited states have a decay rate $\Gamma$, and in the circular basis, the ground states population difference decay rate is $\gamma_{1}$ and the coherence decay rate is $\gamma_{2}$. In our model, when cold atoms are considered, we set $\Gamma/2\pi=6$ MHz, and when a warm atomic vapor is considered, we set $\Gamma/2\pi=500$ MHz to take into account of Doppler broadening.
We consider the steady state of the atoms by setting $\dot{\hat{\rho}}=0$, and then the propagation of the two circular polarized fields are described by the Maxwell equation in the slowly varying amplitude approximation:
\begin{eqnarray}
\frac{\partial \Omega_{1}}{\partial z}=i\kappa(\rho_{23}+\rho_{13})\\
\frac{\partial \Omega_{2}}{\partial z}=i\kappa(\rho_{24}-\rho_{14})
\label{MaxwellEquation}
\end{eqnarray}
where $\kappa=\nu N\wp^{2}/2\epsilon_{0}c\hbar$ is the coupling constant, with $\nu$ the laser frequency, $\wp$ the dipole moment of the transitions $|2\rangle\rightarrow|3\rangle$ and $|2\rangle\rightarrow|4\rangle$ (assumed to be equal), and $N$ the atomic density. Numerically, we can solve for the fields' complex amplitudes $\Omega_{1}(L)$ and $\Omega_{2}(L)$ at the vapor cell output ($L$ the cell length), and then have the probe field's output power as:
\begin{equation}
W_{p}\propto|\Omega_{p}(L)|^{2}=(|\Omega_{1}(L)|^{2}+|\Omega_{2}(L)|^{2}-2|\Omega_{1}(L)||\Omega_{2}(L)|\times\cos{\theta})/2
\label{Interference}
\end{equation}
where $\theta$ is the relative phase between $\Omega_{1}(L)$ and $\Omega_{2}(L)$. It can be seen that the probe field's transmission is related to both the amplitudes and relative phase of $\Omega_{1}$ and $\Omega_{2}$ at the output.

\section{Experimental Setup}

\begin{figure}[b]
  \centering\includegraphics[width=0.5\textwidth]{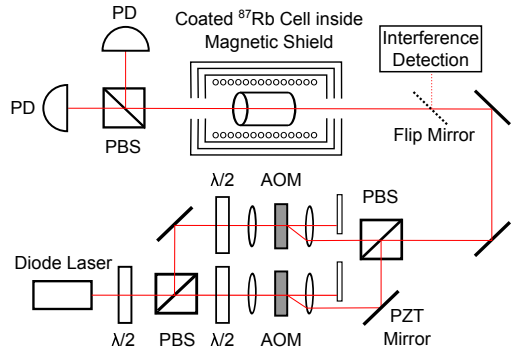}
 \caption{Schematics of the experimental setup. AOM: acoustic-optical modulator, PZT: piezoelectric transducer, PBS: polarization beam splitter,
 PD: photo-detector.}
\label{Setup}
\end{figure}

We performed experiments (schematically shown in Fig.~\ref{Setup}) using a cylindrical paraffin coated vapor cell containing isotropically enriched ${}^{87}$Rb. The cell (2.5 cm diameter, 7.5 cm length) was heated by a blown-air oven residing in a solenoid for magnetic field control, all within a four-layer magnetic shield.  Tuned close to the ${}^{87}$Rb D1 line $F=2$ $\rightarrow$ $F'=1$ transition, the laser passed through a polarized beam splitter (PBS) to generate the linearly polarized pump and the probe field. Then they each went through an acoustic-optical modulator (AOM) for independent power control and pulse shaping. The two identical AOMs shifted the laser frequency by 80 MHz, and the two 80 MHz RF sources were phase locked to ensure the nearly perfect phase coherence between the probe and the control. The first order beams from the AOMs were then combined by another PBS and directed into the vapor cell. At the cell output, a high quality PBS separated the probe and the pump, and they were independently detected by two photo-detectors (PD) with tunable gains.

One of the key elements of this experiment was to have good control on the relative phase between the control and the probe, which was tuned by varying the voltage on the piezoelectric transducer (PZT) attached to the mount of a mirror right before the combining PBS. To reduce fluctuations of the relative phase, the splitting and combining optics were covered by a box to minimize air flow. The maximal phase change the PZT could provide was about $4\pi$. To measure the relative phase at the cell input, we used a flip mirror to direct the beam to a half wave plate followed by a PBS to interfere the control and the probe. Also, it was ensured that this calibration optics at the side gave the same phase with a temporary calibration optics at the cell output (added a half wave plate before the detection PBS) while the entire shield was pushed aside. To check the phase stability, the interference signal was taken before and after each absorption measurement for a particular phase.

In all the experiments reported below, the control field and probe field power were about 225 $\mu$W and 15 $\mu$W respectively (unless otherwise stated), and both of them were 3.5 mm in diameter. We used a single mode fiber at the laser output to ensure good beam profile. The overlap of the two beams after combination were checked by a beam profiler at several locations along the light stream. The temperature of the cell was about 55 ${}^{\circ}\mathrm{C}$, corresponding to an atomic density of about $2\times10^{11}$ cm${}^{-3}$ and an optical depth about 15 \cite{26OptDepth}.

\section{Results and Discussions}
\subsection{Effect of the magnetic field on phase sensitivity}

In this section, we discuss the effect of the magnetic field on the phase sensitivity of the double-lambda system as shown in Fig. 1(b).

\begin{figure}[h]
 \centering\includegraphics[clip,width=0.8\textwidth]{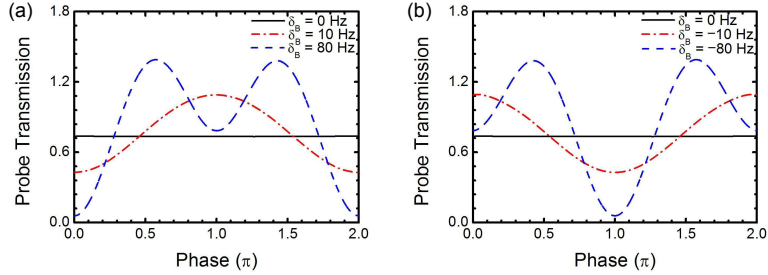}
  \caption{Calculated probe transmission (normalized to the input) vs the relative phase between the probe and the pump for different $\delta_{B}$ in the cold atom regime, where the excited state decay rate $\Gamma/2\pi=6$ MHz. (a) The black solid, red dashdotted and blue dash curves are for $\delta_{B}=0$ Hz, 10 Hz and 80 Hz respectively. (b) The black solid, red dashdotted and blue dash curves are for $\delta_{B}=0$ Hz, $-10$ Hz and $-80$ Hz respectively. Simulation parameters: $\Omega_{c}/2\pi=20$ kHz, $\Omega_{p}/2\pi=5$ kHz, $\gamma_{1}/2\pi=\gamma_{2}/2\pi=10$ Hz, and the optical depth was 0.15. Changes in the optical depth do not affect the main features of the curves as described in the text.}
\label{PhaseSweepColdAtom}
\end{figure}

Let's start from a simpler case which we call the cold atom regime, where the spacing ($\sim800$ MHz) between the two excited states is much larger than the linewidth ($\sim6$ MHz) of the excited states. Fig.~\ref{PhaseSweepColdAtom} shows the calculated probe transmission (normalized to the input value) vs the relative phase between the probe and control, for three magnetic field values (represented by $\delta_B$). Since these curves have a period of $2\pi$, only one period is plotted. It can be seen that, when the magnetic field is off, the probe transmission is insensitive to the phase.  This is because the negligible effect of the upper excited state effectively reduces the system to a normal lambda system, which is not a closed loop without the magnetic field. When the magnetic field is on, the loop is formed. Indeed, when $\delta_B$ increases to 10 Hz and then to 80 Hz, the transmission becomes increasingly sensitive to the phase.
\begin{figure}[h]
 \centering\includegraphics[clip,width=0.8\textwidth]{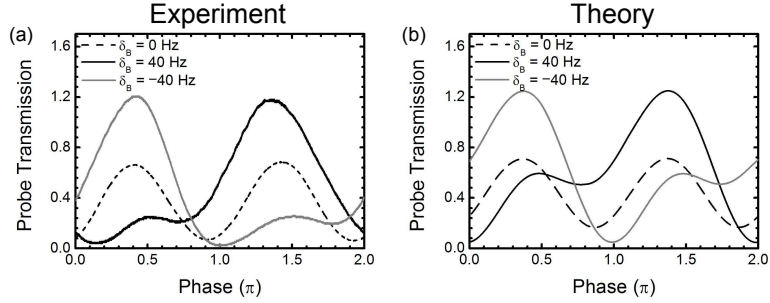}
  \caption{(a) Measured output probe transmission vs the relative phase for different $\delta_{B}$. The black dashed, black solid and grey solid curves are for $\delta_{B}=0$ Hz, 40 Hz and $-40$ Hz respectively. Experimental conditions are in section 3. (b) Corresponding theoretical results where the excited state decay rate $\Gamma/2\pi=500$ MHz. Other simulation parameters: $\gamma_{1}/2\pi=25$ Hz, $\gamma_{2}/2\pi=28$ Hz, $\Omega_{c}/2\pi=240$ kHz, $\Omega_{p}/2\pi=60$ kHz, and the optical depth was 15.}
\label{PhaseSweepWarmAtom}
\end{figure}

Fig.~\ref{PhaseSweepColdAtom}(b) shows the calculated probe transmission for magnetic fields opposite to that in Fig.~\ref{PhaseSweepColdAtom}(a). It was found that two curves with opposite $\delta_B$ values coincide if we translate one of them by $\pi$ on the x-axis. This phenomenon can be explained using the linear basis (Fig.~\ref{FWMpic}(b)). In each lambda loop, for example, in the cycle of $|2\rangle\rightarrow|X\rangle\rightarrow|Y\rangle\rightarrow|2\rangle$, the coherence term $\rho_{2Y}$ is determined by the product of $\Omega_{c}$ and $\delta_{B}$ which remains unchanged if the sign of $\delta_{B}$ and $\Omega_{c}$ reverses simultaneously. Indeed, we found that the master equation remains valid when the following substitution is made: $\delta_{B}\rightarrow-\delta_{B}$, $\Omega_{p}\rightarrow\Omega_{p}$, $\Omega_{c}\rightarrow-\Omega_{c}$, $\rho_{1X}\rightarrow\rho_{1X}$, $\rho_{2Y}\rightarrow\rho_{2Y}$, $\rho_{1Y}\rightarrow-\rho_{1Y}$, $\rho_{2X}\rightarrow-\rho_{2X}$, $\rho_{XY}\rightarrow-\rho_{XY}$, $\rho_{aa}\rightarrow\rho_{aa}$ ($a=1, 2, X, Y$), which indicates these are two sets of identical solutions to the system. In other words, the probe field susceptibility should be the same if we change the sign of the magnetic field and change the relative phase from $\varphi$ to $\varphi+\pi$ simultaneously.


To verify above predictions, we measured the probe transmission vs the relative phase for $\delta_{B}=0$, $\pm40$ Hz (Fig.~\ref{PhaseSweepWarmAtom}). The transmission was normalized to the off-resonant transmission of the probe when the control was absent. We can see that, in contrast to the cold atom regime, the phase sensitivity was already very pronounced without the magnetic field. This is because that Doppler broadening makes the effects of the upper excited state not negligible, and thus the fields form a closed loop four-wave-mixing (even without the magnetic field) which is phase sensitive. When $\delta_{B}$ was nonzero, the phase sensitivity increased, as in the simulation for cold atoms. Also, the two curves for $\delta_{B}=\pm40$ Hz can almost overlap when one curve is shifted by $\pi$ along the x-axis, consistent with above theoretical analysis. To account for the shape difference of the measured curves compared to the cold atom case, we performed simulations with $\Gamma/2\pi=500$ MHz (Fig.~\ref{PhaseSweepWarmAtom}(b)) and found good agreement with our experiment. The experiment results show that a double-lambda scheme has distinct phase dependence than a single lambda scheme.

\subsection{System manipulation via combination of relative phase and magnetic field}

\begin{figure}[ht]
 \centering\includegraphics[clip,width=0.8\textwidth]{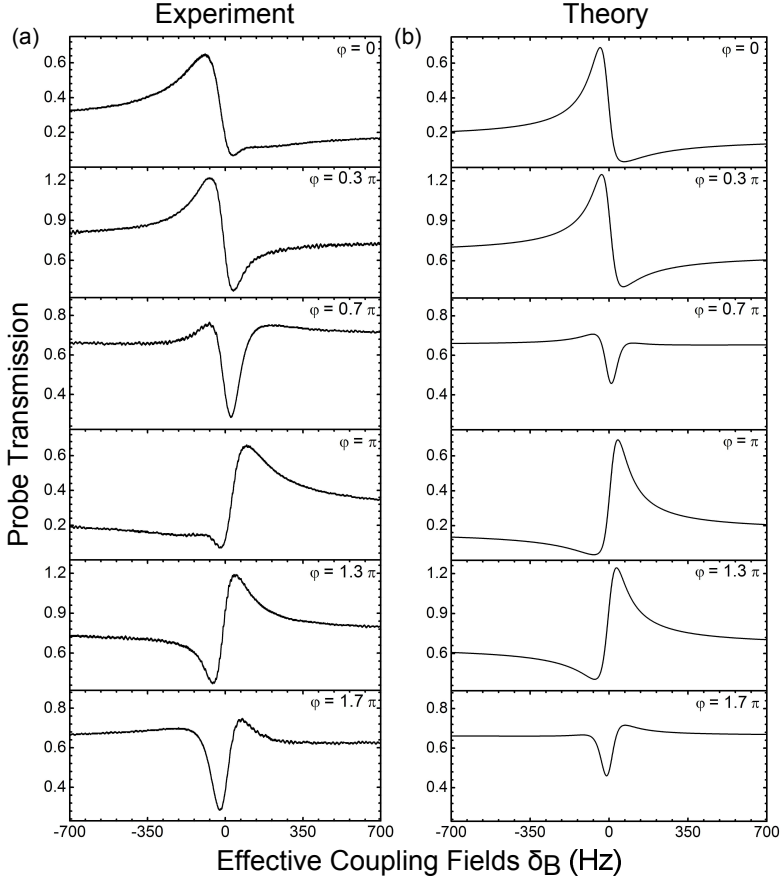}
  \caption{Probe transmission vs $\delta_{B}$ for several relative phases. (a) Measured results, with experimental conditions stated in section 3. (b) Calculated results, with identical simulation parameters as used in Fig.~\ref{PhaseSweepWarmAtom}(b).}
\label{55}
\end{figure}
In this section, we investigate the combined effects of the magnetic field and the relative phase on the probe's transmission and its group velocity.

We first measured the probe transmission vs $\delta_{B}$ for various relative phase $\varphi$. As shown in Fig.~\ref{55}(a), the curves combine features of absorption and dispersion plots, which can be interpreted using the circular basis. Eq. (6) shows that the probe transmission carries the information of both the absorption and refractive index (phase) of the circular laser beams. In particular, the first two terms of Eq. (6) represent the total transmission of the two circular fields, and should give a normal Lorentzian EIT spectra when $\delta_B$ is swept. As for the interference term of Eq. (6), $\cos{\theta}$ can be written as $\cos{\theta}=\cos{(\theta_0+\varepsilon)}=\cos{\theta_0}-\varepsilon\sin(\theta_0)$ if $\varepsilon$ is small. Here $\theta_0$ is the input relative phase between the two circular beams, and $\varepsilon$ is the accumulated phase difference which has a dispersive feature.

Also, it is clearly seen from Fig.~\ref{55} that, the two curves with relative phase $\varphi$ and $\varphi+\pi$ are mirror images of each other. This again verifies that the probe transmission remains unchanged if we change the sign of $\delta_{B}$ and add $\pi$ to the relative phase simultaneously. Fig.~\ref{55}(b) is the numerically calculated results, which agree well with the experiment. Remaining discrepancy between the experiment and the simulation results is mainly due to that in our simplified model, we did not include the effects of atomic motion in the coated vapor cell \cite{27CoatedCell,28CoatedCell,29CoatedCell}, and Doppler broadening was not strictly taken into account by integrating over all velocities.

\begin{figure}[t]
 \centering\includegraphics[clip,width=0.8\textwidth]{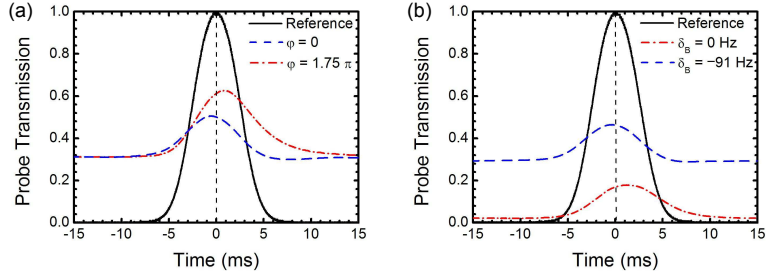}
  \caption{Measured slow and fast light results. Probe pulse transmission is normalized to the reference (black solid curve) pulse. The blue dashed curve is the fast light, and the red dashdotted curve is the slow light. (a) $\delta_{B}=-91$ Hz. (b) relative phase $\varphi=0$.}
\label{FastSlowLight}
\end{figure}

Next, we studied the dynamic behavior of the probe field, and found that the refractive index and the group velocity can be also controlled by the magnetic field and the relative phase. We generated the probe pulse field via programming the AOM's driving field by LabVIEW. For optimal slow and fast light effects, the full width half maximum (FWHM) of the probe pulse was chosen to be about 5 ms. The continuous control field power was 225 $\mu$W and the peak power of the probe pulse was 15 $\mu$W. When $\delta_{B}$ was set at $-91$ Hz, we could turn the slow light into fast light by changing the relative phase from about 1.75 $\pi$ to 0 (Fig.~\ref{FastSlowLight}(a)). The base of the fast light or slow light curve was the result of the optical rotation of the control field due to the magnetic field. On the other hand, if the relative phase was set to be 0, we could also turn the slow light into fast light by changing $\delta_{B}$ from 0 Hz to $-91$ Hz (Fig.~\ref{FastSlowLight}(b)). The fractional delay and advance were about 25$\%$ and 10$\%$ respectively.

\begin{figure}
 \centering\includegraphics[clip,width=0.8\textwidth]{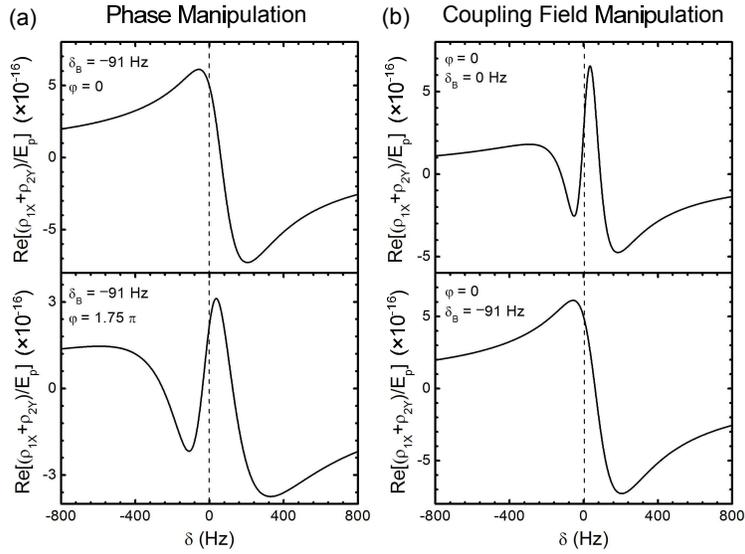}
  \caption{Calculated spectrum of the real part of $(\rho_{1X}+\rho_{2Y})/E_p$ vs the probe frequency, with the pump frequency fixed. (a) Relative phase manipulation for $\delta_{B}=-91$ Hz. (b) Magnetic field manipulation for the relative phase $\varphi=0$. Simulation parameters are the same as in Fig.~\ref{PhaseSweepWarmAtom}(b).}
\label{RefractiveIndex}
\end{figure}

To explain such phenomena, we calculated the refractive index experienced by the probe field while fixing the control field frequency and sweeping the probe's one-photon detuning $\delta$. For this calculation, we rewrote the Hamiltonian of the system in the linear basis as depicted in Fig.~\ref{FWMpic}(b), and computed the real part of $(\rho_{1X}+\rho_{2Y})/E_p$ which is proportional to the refractive index for the probe. According to the group velocity formula
\begin{equation}
v_{g}=\frac{c}{n+\omega\frac{{\rm d}n}{{\rm d}\omega}},
\end{equation}
the slope of the refractive index spectra of the probe near $\delta=0$ Hz in Fig.~\ref{RefractiveIndex} determines the group velocity of the probe pulse. Fig.~\ref{RefractiveIndex}(a) shows that when $\delta_{B}=-91$ Hz, the slope near $\delta=0$ changes from positive to negative when the phase changes from 1.75 $\pi$ to 0. Fig.~\ref{RefractiveIndex}(b) shows that when the relative phase $\varphi=0$, the slope near $\delta=0$ changes from positive to negative when $\delta_{B}$ varies from 0 Hz to $-91$ Hz. This verifies that switching between slow and fast light can be realized by either adjusting the magnetic field or the relative phase.

\section{Conclusion}
In conclusion, we demonstrated a new method to manipulate the phase sensitivity of a double-lambda system. A magnetic field coupling the two ground states can function as an effective oscillating electromagnetic field, and thus can either close an otherwise open interaction loop, rendering the system phase-sensitive, or alter the phase sensitivity of an initially phase-dependent system. We have theoretically and experimentally verified that a static magnetic field can drastically influence the absorption and refractive index of an optical field and its dependence on the relative phase.  This work should be useful for group velocity manipulation, nonlinear optics and magnetometery.

\section*{Acknowledgments}
We are grateful to Xiangming Hu and Lei Feng for useful discussions. This work was supported by the NBRPC (973 Program Grants No. 2011CB921604 and No. 2012CB921604), the NNSFC (Grant No. 61078013), and Fudan University.
\end{document}